# EUKulele: Taxonomic annotation of the unsung eukaryotic microbes

**Arianna I. Krinos**[1,2], **Sarah K. Hu**[3,4], **Natalie R. Cohen**[3], **and Harriet Alexander**[*,1]

**1** Biology Department, Woods Hole Oceanographic Institution **2** MIT-WHOI Joint Program in Oceanography **3** Marine Chemistry and Geochemistry, Woods Hole Oceanographic Institution **4** Center for Dark Energy Biosphere Investigations

## Summary

The assessment of microbial species biodiversity is essential in ecology and evolutionary biology (Reaka-Kudla, Wilson, & Wilson, 1996), but especially challenging for communities of microorganisms found in the environment (Das, Lyla, & Khan, 2006; Hillebrand et al., 2018). Beyond providing a census of organisms in the ocean, assessing marine microbial biodiversity can reveal how microbes respond to environmental change (Salazar & Sunagawa, 2017), clarify the ecological roles of community members (Hehemann et al., 2016), and lead to biotechnology discoveries (Das et al., 2006). Computational approaches to characterize taxonomic diversity and phylogeny based on the quality of available data for environmental sequence datasets is fundamental for advancing our understanding of the role of these organisms in the environment. Even more pressing is the need for comprehensive and consistent methods to assign taxonomy to environmentally-relevant microbial eukaryotes. Here, we present EUKulele, an open-source software tool designed to assign taxonomy to microeukaryotes detected in meta-omic samples, and complement analysis approaches in other domains by accommodating assembly output and providing concrete metrics reporting the taxonomic completeness of each sample.

EUKulele is motivated by ongoing efforts in our community to create and curate databases of genetic and genomic information (Allen, 2015; Caron et al., 2017; UGA, 2020). For decades, it has been recognized that genetic and genomic techniques are key to understanding microbial diversity (Fell, Statzell-Tallman, Lutz, & Kurtzman, 1992). Genetic approaches are particularly useful in poorly-understood or difficult-to-access environmental systems, which may have a high degree of species diversity (Das et al., 2006; Mock et al., 2016). The most common approach for censusing microbial diversity is genetic barcoding, which targets the hyper-variable regions of highly conserved genes such as 16S or 18S rRNA (Leray & Knowlton, 2016). Computational approaches to assess the origin of these barcode-based studies (or tag-sequencing) have been well established (Bolyen et al., 2018; Schloss et al., 2009), and enable biologists to compare microbial communities and estimate sequence phylogeny. The recent collation of reference databases, e.g. PR2 and EukRef, for ribosomal RNA in eukaryotes have enabled more accurate taxonomic assessment (Del Campo et al., 2018; Guillou et al., 2012). However, barcoding approaches that focus on single marker genes or variable regions limit the field of view of microbes–especially protists, which have complex and highly variable genomes (Campo et al., 2014)–potentially limiting the organisms recovered and leaving the "true" diversity poorly constrained (Caron & Hu, 2019; Piganeau, Eyre-Walker, Grimsley, & Moreau, 2011).

*Corresponding author

Shotgun sequencing approaches (e.g. metagenomics and metatranscriptomics) have become increasingly tractable, emerging as a viable, untargeted means to simultaneously assess community diversity and function. Large-scale meta-omic surveys, such as the Tara Oceans project (Zhang & Ning, 2015), have presented opportunities to assemble and annotate full "genomes" from environmental metagenomic samples (Delmont et al., 2018; Tully, Graham, & Heidelberg, 2018) and assemble massive eukaryotic gene catalogs from environmental metatranscriptomic samples (Carradec et al., 2018). The interpretation of these meta-omic surveys hinges upon curated, culture-based reference material. Several such curated databases that contain predicted proteins from a mixture of genomic and transcriptomic references from eukaryotes, as well as bacteria and archaea, have been created (e.g. PhyloDB (Allen, 2015), EUKZoo (Liu, 2018), MarineRefII (UGA, 2020), and EukProt (D. J. Richter et al., 2020)). The Marine Microbial Eukaryote Transcriptome Sequencing Project (MMETSP) database, which contains over 650 fully-assembled reference transcriptomes (Johnson, Alexander, & Brown, 2019; Keeling et al., 2014), is among the largest single projects to create a unified reference. These databases are an invaluable resource yet, to our knowledge, no single integrated software tool currently exists to enable an end-user to harness databases in a consistent and reproducible manner. Tools like MEGAN (Beier, Tappu, & Huson, 2017) and MG-RAST (Keegan, Glass, & Meyer, 2016) have been developed for annotation and analysis of metagenomes and extended to metatranscriptomes, but they are optimized for performance with NCBI databases, which might not be well-curated for the study of environmental eukaryotes. Further, existing tools do not provide a single function that can download and format databases, which is necessary for computational tools to remain relevant and usable as reference databases grow.

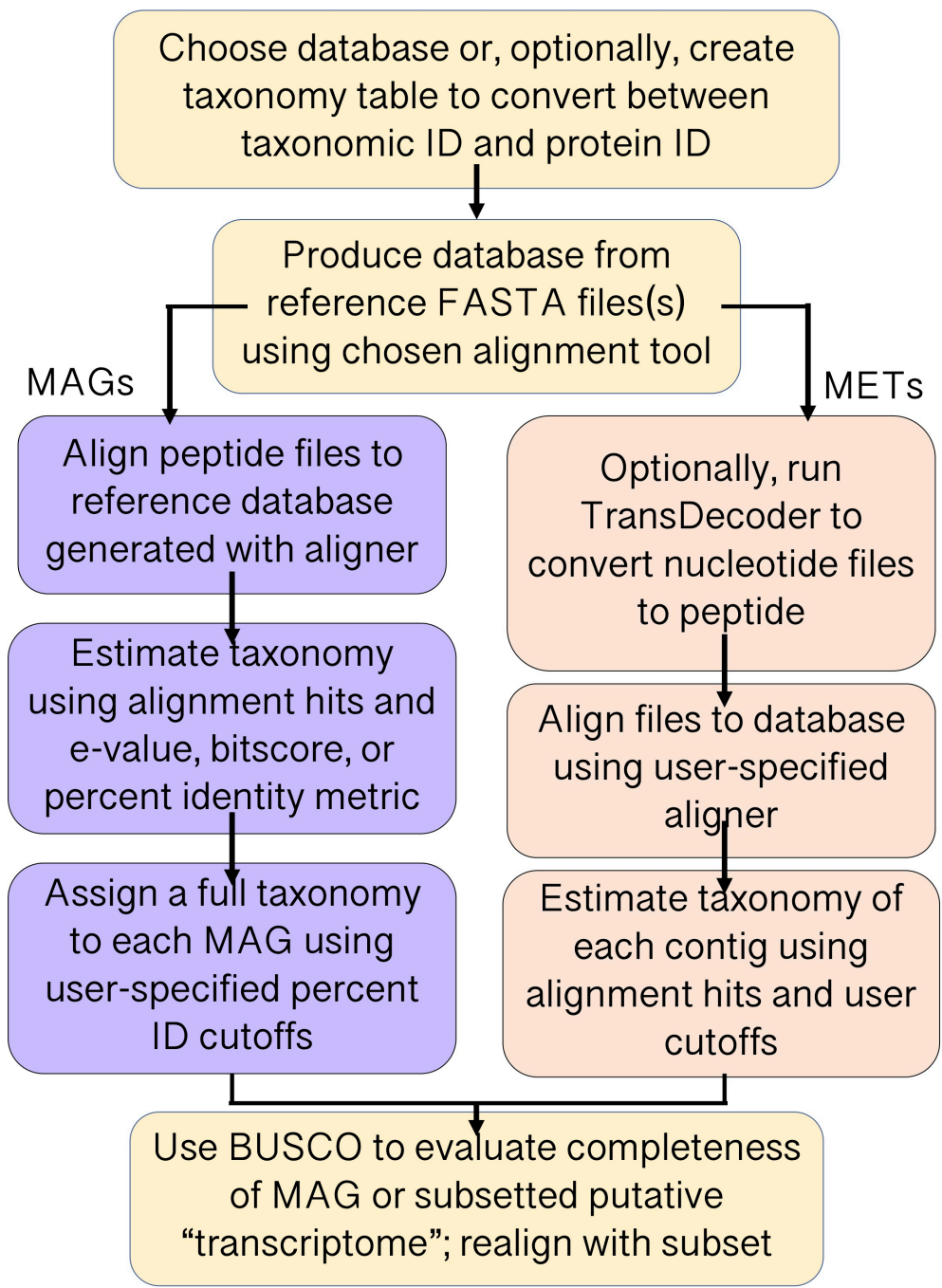

**Figure 1:** A flowchart describing the general workflow of the software as it relates to metatranscriptomes (METs) and metagenomes (MAGs).

`EUKulele` (Krinos, Hu, Cohen, & Alexander, 2020) (Figure 1) is an open-source `Python`-based package designed to simplify taxonomic identification of marine eukaryotes in metaomic samples. User-provided metatranscriptomic or metagenomic samples are aligned against a database of the user's choosing, using a user-chosen aligner (`BLAST` (Kent, 2002) or `DIAMOND` (Buchfink, Xie, & Huson, 2015)). The "blastx" utility is used by default if metatranscriptomic samples are only provided in nucleotide format, while the "blastp" utility is used for samples available as translated protein sequences. Any consistently-formatted database may be used, but three published microbial eukaryotic database options are provided by default: MMETSP (Caron et al., 2017; Johnson et al., 2019; Keeling et al., 2014), PhyloDB (Allen, 2015), and EukProt (Richter et al., 2020). The package returns comma-separated files containing all of the contig matches from the metatranscriptome or metagenome, as well as the total number of transcripts that matched, at each taxonomic level, from supergroup to species. If a quantification tool has been used to estimate the number of counts associated with each transcript ID, counts may also be returned.

### Statement of Need

A growing number of databases have been created to catalog eukaryotic and bacterial diversity, but even when the same database is used, taxonomic assessment is not always consistent and fully documented (Menzel, Ng, & Krogh, 2016; Rasheed & Rangwala, 2012). Databases often contain distinct compilations of organisms and custom databases are commonly compiled for only a particular study (Geisen et al., 2015; Kranzler et al., 2019; Obiol et al., 2020). Database variability might influence interpretation by splitting taxonomic annotations between groups, and often impacts the proportion of contigs that are annotated (Price & Bhattacharya, 2017). A software tool can bridge the gap between database availability and efficient taxonomic assessment, making environmental meta-omic analyses more reproducible. Further, such a tool can control and assess the quality of the annotation, enable inference for specific organisms or taxonomic groups, and provide more conservative annotation in the case of organisms with exceptional amounts of inter-strain variability. We have designed the `EUKulele` (Krinos et al., 2020) package to enable efficient and consistent taxonomic annotation of metagenomes and metatranscriptomes, in particular for eukaryote-dominated samples.

### Future Outlook

As single species isolates continue to be sequenced, databases are growing and becoming more reliable for assigning taxonomy in diverse environmental communities. `EUKulele` provides a platform to enable the repeated and consistent linkage of these databases to metagenomic and metatranscriptomic analyses. Taxonomic annotation is not the only desired outcome of meta-omic datasets against available databases, hence we envision eventually integrating functional annotation into the `EUKulele` package.

## Acknowledgements

This software was developed with support from the Computational Science Graduate Fellowship (DOE; DE-SC0020347) awarded to AIK and from the Woods Hole Oceanographic Independent Research & Development grant awarded to HA. NRC was supported by grant 544236 from the Simons Foundation. The Center for Dark Energy Biosphere Investigations (C-DEBI; OCE-0939564) supported the participation of SKH through a C-DEBI Postdoctoral Fellowship. The High Performance Computing cluster at Woods Hole Oceanographic Institution (Poseidon) was used to generate assemblies and run `EUKulele`.

## Author Contributions

HA and SKH conceived the original idea for the tool. HA wrote the initial code for the tool. HA and AIK refined ideas for `EUKulele` related to designing it as an installable package and adding the feature of classification via core gene taxonomy for metagenomic applications. AIK developed the Python package code, wrote the conda package, implemented multiple alignment tools, and the `BUSCO` integration. AIK, NC, SKH, and HA wrote tests and documentation. HA and AIK wrote the paper.